# A Sliding Ferroelectric Resonant Tunnel Junction

Noam Raab[1,6] Renu Yadav,[2,6] Yakov Bloch[2], Youngki Yeo[2], Chen Maoz[2], Iva Plutnarova[3], Zdenek Sofer[3], Watanabe Kenji[4], Takashi Taniguchi[5], Moshe Ben Shalom[1,#]

[1]SlideTro Ltd, Tel Aviv, Israel

[2]School of Physics and Astronomy, Tel Aviv University, Tel Aviv, Israel

[3]Dept. of Inorganic Chemistry, University of Chemistry and Technology Prague, Technicka 5, 166 28 Prague 6, Czech Republic

[4]Research Center for Functional Materials, National Institute for Materials Science, Tsukuba, Japan

[5]International Center for Materials Nanoarchitectonics, National Institute for Materials Science, Tsukuba, Japan

[6]These authors contributed equally: Noam Raab and Renu Yadav

[#]Corresponding author: Moshe Ben Shalom (moshebs@tauex.tau.ac.il)

Ferroelectric tunnel junctions (FTJs) leverage polarization-dependent tunneling through ultrathin barriers to enable two-terminal, non-volatile memory and logic[1–3]. Although conceptually appealing, the practical implementation of conventional FTJs has been hindered by high coercive voltages, low readout currents, limited cycling endurance and significant device-to-device variability[4–7]. Here, we overcome these bottlenecks by introducing the sliding ferroelectric resonant tunnel (SFeRT) junction, integrating three cooperative mechanisms: (i) spontaneous interfacial polarization of atomically thin[8–11], depolarization-resilient[12,13] barriers; (ii) superlubric sliding of shear-solitons [10,14,15], enabling ultra-low-friction[16,17], wear-free switching [18,19]; and (iii) momentum-conserving, elastic resonant tunneling between lattice-aligned graphitic electrodes[20,21], providing sensitive readouts at both positive and negative biases.

We demonstrate nanometer-scale SFeRT junctions using polar polytypes of hexagonal boron nitride ($h$BN) or transition metal dichalcogenides (TMDs) as barriers, achieving configurable writing voltages below 0.5 V and tunable reading biases under 0.1 V. These devices yield current densities exceeding 50 nA µm$^{-2}$, with a robust room-temperature ON/OFF ratio > 7.

The crystalline and polarization integrity of sliding van der Waals (vdW) polytypes [15], down to the atomically thin limit, ensures exceptional device uniformity and performance that remains scalable down to sub-0.1 µm$^2$ footprints. Furthermore, we provide a predictive model for SFeRT performance across diverse doping levels, temperatures, electrodes, and polytype configurations. Integrated within a Superlubric Array of Polytypes (SLAP) [16] architecture, SFeRT junctions enable switching energies below 1 fJ, [17] establishing a scalable and durable foundation for low-energy "slidetronic" logic and memory.

**Main**: The probability of an electron to tunnel across energy barriers is determined by the spatial overlap of orbital wavefunctions, which decays exponentially with increasing separation - typically by four orders of magnitude per nanometer. For instance, a 0.1 Volt bias ($V_{ds}$) between a drain and a source electrode separated by $h$BN yields a tunneling current density ($J$) that falls by a factor of 20 for each additional atomic layer, from $10^2$ µA/µm² in monolayer $h$BN spacer (0.34 nm thick) to $10^{-3}$ µA/µm² in four-layers barriers (1.38 nm thick)[22]. Such a rapid attenuation of short-range orbital overlap imposes rigorous constraints on conventional ferroelectric tunnel barriers, including ferroelectric perovskites and HfO$_2$-based systems. At this ultimate thin-film limit, the emergence of potent depolarization fields, structural inhomogeneities, and accelerated switching fatigue presents a formidable barrier to the practical realization of high-performance FTJs [2–7].

Layered vdW heterostructures provide a robust solution to the atomically thin-crystal limit challenges, preserving structural and electronic integrity down to monolayers [23]. Their pristine, dangling-bond-free surfaces suppress detrimental substrate interactions, while strong in-plane covalent hybridization ensures exceptional stiffness and mechanical stability. Although the chemical inertness of vdW interfaces might suggest negligible interlayer interactions, a spontaneous room-temperature out-of-plane polarization ($P$) emerges[24], enhanced by commensurate atomic registries and the discrete symmetries of vdW polytypes [15]. Specifically, parallel bilayers of honeycomb binary compounds (e.g., $h$BN or WSe$_2$ TMD) exhibit spontaneous interlayer charge transfer of ≈ 5 × 10¹² $e$/cm² due to the breaking of inversion and mirror symmetries[8–13,25,26].

In these AB-stacked configurations (Fig. 1a), the A-site atoms of one layer eclipse the B-site atoms of the successive layer, generating a built-in potential of $V_P$≈ 0.12 V ($h$BN), 0.06 V (WSe$_2$) per interface. Notably, increasing the layer count leads to a linear potential accumulation[12] up to the band gap energy (5.9 eV and 1.4 eV, respectively)[13], confirming the robustness against depolarization in the thin-layer limit. Within a bilayer framework, each interlayer shift between the two stable AB and BA configurations is equivalent to flipping the structure (and the polarization $P$) upside down[8]. Furthermore, in trilayer stacks and beyond, these shifts modify the material type, providing additional multiferroic responses[27,28]. For example, a stack of four parallel $h$BN layers can transition between three metastable structural phases - so-called vdW polytypes[15] - ABCA, ABAB, ABCB, yielding a manifold of four (±$V_P$, ±3$V_P$) discrete, evenly spaced, non-volatile potential states[12,13].

The switching dynamics of SFeRT junctions also deviates fundamentally from conventional FTJs. Rather than out-of-plane ionic displacement[2], electric field switching in vdW polytypes is driven by planar sliding of a pre-existing boundary wall[10] - incommensurate shear soliton strip - that carries a single bond-length shift between the layers (a topological partial-dislocation)[29]. As a shear soliton traverses the structure, the particular interface switches between two stable stacking configuration accompanied by a sharp 2$V_P$ potential jump[15]. Notably, uniform polytype structures without solitons remain resilient to electric switching due to a *C3* symmetry, which forbids off-diagonal coupling between vertical electric fields and in-plane atomic displacement[10,14,15,30–33]. The combination of high in-plane stiffness (TPa range) and a shallow commensurate potential well of only ≈ 1 meV/atom extends the vdW soliton width to over ≈ 30 atoms[34], raising the energy of minimal structural excitations beyond 1 eV (for 30×30 incommensurate atoms) and ensuring stability ($T_c > 1000$ K)[15,35]. Moreover, these sliding solitons are largely decoupled from underlying phonons and local disorder-induced pinning, resulting in superlubric, ballistic-like sliding and exceptionally low coercive fields[10,36–38]. This stability-efficiency trade-off is further optimized via the Superlubric Array of Polytypes (SLAP) architecture[16]. By embedding active parallel layers within cavities etched into a misaligned vdW spacer, SLAP exploits a four-order-of-magnitude reduction in the friction coefficient at the incommensurate spacer surfaces[39,40]. This architecture facilitates reversible switching of sub-50 nm domains with shape-tuneable switching energies below 1 fJ and enables novel elastic coupling of the coercive fields in adjacent cavities via long-range strain relaxation[41].

Finally, readout in FTJs is characterized by the tunneling electroresistance (TER), defined as the ratio between high- and low-resistance states at a specific $V_{ds}$. While conventional FTJs are optimized to supress the density of states (DOS) at a single electrode and *P* OFF state [3–6], SFeRT junctions utilize identical, lattice-aligned graphitic electrodes[20,42–44] (illustrated in Fig. 1a) that maintain a substantial DOS across all available polarization states. Rather than modulating the DOS, the applied bias in a SFeRT junction triggers momentum-conserving tunneling by aligning perfectly matched energy bands in the ON-state - a condition that is deactivated in alternative polarization configurations. As demonstrated, this elastic resonant tunneling mechanism enables current densities 100-fold (room temperature) to 1,000-fold (cryogenic temperature) higher than the inelastic scattering mechanisms dominant in conventional FTJs [45,46]. It facilitates continuously tunable ultra-low $V_{ds}$ reading and unique gate-tunable writing capabilities.

**SFeRT devices**

A schematic cross-section of a typical SFeRT junction is shown in Figure 1a. As a prototype sliding ferroelectric barrier, we assemble 4-layer thick polytypes of *h*BN (Fig.1) or $WSe_2$ (Fig.2) featuring a middle polar (AB) interface by stacking a parallel pair of anti-aligned A'A bilayers (where the prime denotes a 180-degree twist). We use a pair of bilayer graphene (BLG) flakes as source and drain electrodes that are co-aligned with each other but remain incommensurate with the barrier surfaces due to lattice mismatch and unintentional twist misalignment (Fig. S1). A gate electrode facing the BLG source is used to tune its charge density by controlling the gate bias $V_{gs}$. In the supplementary information (SI), we discuss junctions incorporating other polytypes, demonstrating the structural and polar versatility of the SFeRT architecture. To shrink the junction area below 0.1 µm² and obtain a uniform single-domain switching, we insert a ≈ 2 nm thick *h*BN dielectric spacer with etched holes allowing finite tunneling currents at the holes' position only[41] (Fig.1a, b). Alternatively, we study large multi-domain junctions, without a confining cavity, to probe the current through both polarizations in parallel (Fig. 2a), providing a clear comparison of the underlying temperature and gate-dependent physics.

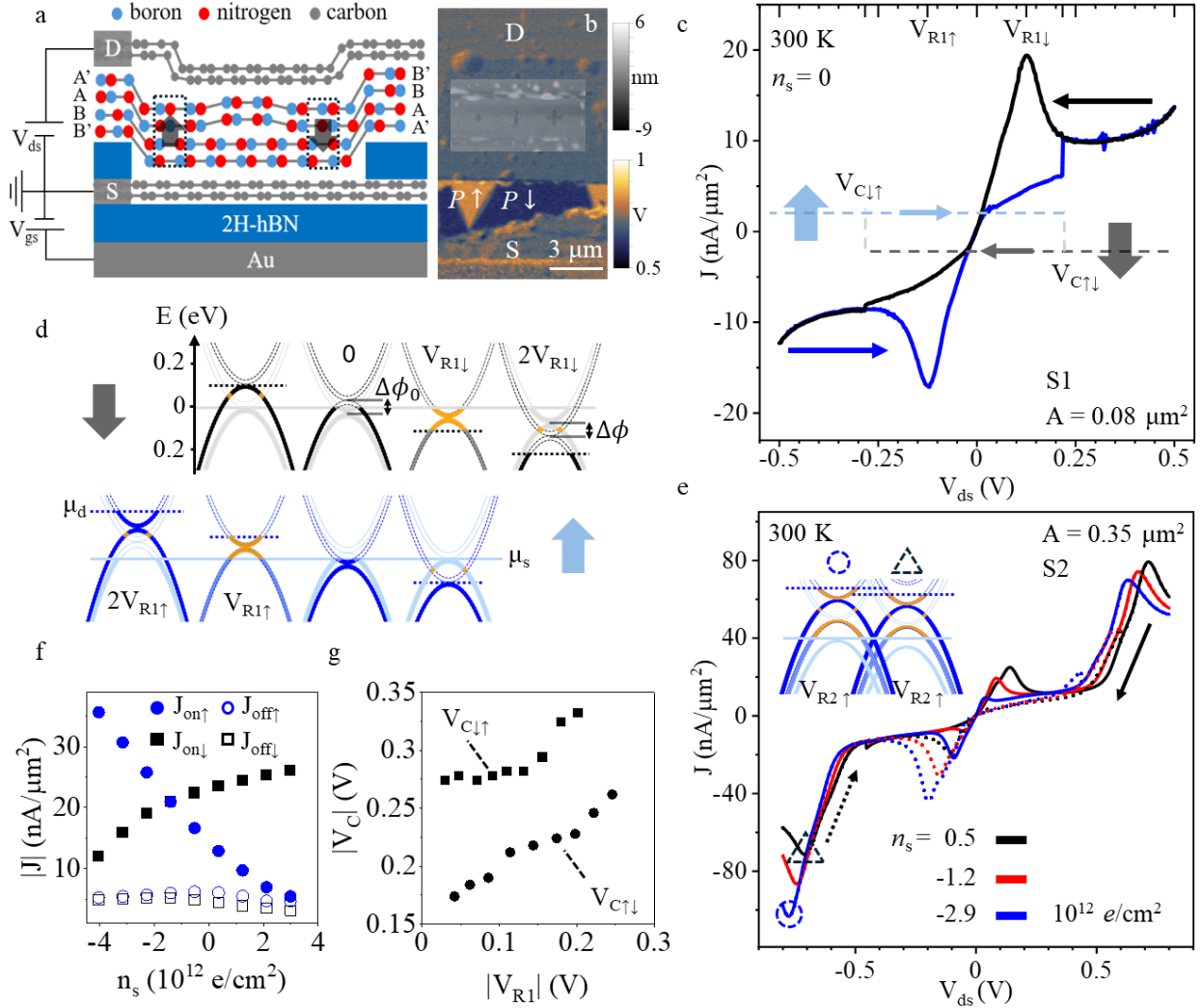

*Figure 1: SFeRT structure, band alignments, and current-voltage switching response.* (a) Cross-section illustration of a junction and the measurement circuit diagram. The shear soliton slides between two sides of the junction to switch the local interfacial polarization (b) AFM topography (grayscale) map overlayed on a surface potential (orange-navy) map of the device showing polar domains and cavities. Polar domains can be detected only in the region where top BLG is not present to screen them. (c) Hysteretic current density response versus drain-source bias ($J$-$V_{ds}$) of the SFeRT device shown in b. The blue (black) curve corresponds to a forward (backward) sweep. Horizontal dashed lines indicate the up and down polarization states. (d) Band structure diagrams at characteristic $V_{ds}$ values for up and down polarization in blue and black, respectively. Source (drain) bands are in solid light (dashed dark) color. Active resonant tunneling overlaps are highlighted in orange. (e) $J$-$V_{ds}$ response for various charge densities at the source $n_s$, and fixed gate values $V_{gs}$. Band structure diagrams of the second resonance peak $V_{R2}$ at two doping levels. (f) ON and OFF current densities of both ↑,↓ polarization orientations for various $n_s$ levels. (g) Coercive switching voltage $V_c$ vs. the peak position $V_{R1}$.

**Tunneling and switching single-domain *h*BN junctions:**

Figure 1c displays a hysteretic tunneling current density ($J$) as a function of $V_{ds}$ for a typical 0.08 µm² single-domain junction. The blue and black curves are measured during a forward and backward scan, respectively, showing an abrupt polarization switch at $V_c$ = 0.22, -0.28 V (marked by dashed vertical lines), where $J$ jumps between the two curves at a single $V_{ds}$ point. The $J_{ON\uparrow\downarrow} \approx \pm 18$ nA/µm² resonant peaks at $V_{R1} = \pm 0.123$ V mark conditions at which the drain and source BLG bands align in energy, see the solid and dashed bands respectively in the $\pm V_{R1}$ diagrams, panel d, for $P\downarrow$ (black) and $P\uparrow$ (blue) states. At this first resonance, an electrostatic potential difference $\Delta\phi_0$ (band shift) generated by the spontaneous intrinsic polarization at zero-bias (see $V_{ds} = 0$ diagrams, panel d) - is countered by an opposite bias: $V_{R1} = -V_P$. This compensation opens a momentum-matched (see the bands' overlaps coloured in orange) electrochemical energy window $\Delta\mu_R = eV_{R1}$ (dashed to solid horizontal lines), within which, filled source states tunnel into empty drain states. Conversely, in the OFF state with the same $V_{ds}$ and opposite $P$ orientation, the electrostatic bands shift doubles to $-2\Delta\phi_0$, leaving only two band-crossing points (orange points) rather than a fully overlapping energy window, resulting in a suppressed current $J_{OFF} \approx 4$ nA/µm². The measured $J_{ON}/J_{OFF} \approx 5$ resembles a $\Delta\mu_R/k_BT$ ratio, as expected for a room-temperature response dominated by thermal broadening.

Scanning $V_{ds}$ to $2V_{R1}$ pushes the bands beyond the resonant condition (see the $2V_{R1\uparrow,\downarrow}$ diagrams), resulting in a negative differential conductance[20] (NDC). Here, the smaller peak-to-valley factor ($\approx 3$ rather than 5) arises from finite inelastic and thermally excited tunneling[5] as discussed below. Movie 1 (SI) presents detailed calculations of the bands' overlaps and occupations for all $V_{ds}$ points, incorporating the $h$BNs' geometric capacitance, a depolarization factor - $P(V_{ds})$, broadening due to inelastic scattering ($\Gamma$) and temperature (T), and the BLGs' quantum capacitance ($C_q$), trigonal wrapping, and finite twist angle $\theta$.

**Band alignment picture**

To follow the evolution of $J_{ON}$ and $J_{OFF}$, we first consider a simplified picture of overlapping energy-momentum paraboloids and rings (orange parabolas and points in the cross-sectional diagrams), neglecting depolarization. We identify second resonant peaks appearing at $V_{R2} = \pm 0.7$ V with $J_{ON2} = \pm 80$ nA/µm² (panel e, dashed triangle). These peaks correspond to an electrostatic shift of $\Delta\phi = \gamma_1 = 0.39$ eV (the BLG binding energy, top diagram) between the source and drain bands towards each other to overlap a deeper band of one with the first band of the other. [47]

A $\gamma_1$ shift that requires a bias shift of $e\Delta V_{ds} = 0.58$ $eV$ between $V_{R1}$ and $V_{R2}$, indicates a $\frac{0.39}{0.58} = 0.67 \approx \frac{\Delta\phi}{e\Delta V_{ds}}$ ratio of electrostatic to electrochemical shift, which remains nearly constant in this range (up to depolarization effects) due to the fixed DOS in BLG. This ratio confirms the expected response for a $C_q/A \approx 2.2$ µF/cm² of the BLG pair and a $C_g/A \approx 2.3$ µF/cm² of a 4-layer hBN spacer.

Since $J_{ON}$ is determined by the number overlaping states within the $\Delta\mu_R = eV_R$ window, it can be modulated either by the spontaneous polarization-induced voltage $V_P$, or by tuning the gate voltage $V_{gs}$ and the zero-bias charge density at the source $n_s$. A negative $V_{gs}$, for example, pulls electrons from the source, increasing (decreasing) the bands-missalignment for the $P\downarrow$ ($P\uparrow$) state, thereby shifting $V_{R1,2}$ away from (towards) zero bias. This reasoning explains the linear $V_{gs}$ - tuning of $|V_{R1}|$ and $|J_{ON1}|$ in panel e (dashed curves) with $n_s = 0.5, -1.2, -3 \times 10^{12}$ $e$/cm² and $J_{ON\uparrow} =$ $-19, -31, -44$ nA/μm² for $V_{R\uparrow} = -0.11, -0.15, -0.2$ V, respectively. Conversely, $J_{OFF}$ remains nearly constant, independent of the bias window, since the number of states at the (orange) band-crossing points is energy-independent as long as they lie within $\Delta\mu_R$. Together, these effects tune the room temperature TER from 5 at $n_s = 0$, to 8 at $n_s = 3.5 \times 10^{12}$ $e$/cm² (panel f).

At the second resonant, an even sharper gate response emerges (see the dashed black triangular and blue circular marks, panel e). Here, shifting $V_{R2}$ away from zero bias opens the electrochemical window $\Delta\mu_R$ while simultaneously increasing the fraction of overlapping states within it. Unlike the complete overlap at $V_{R1}$, only a subset of states in this $V_{R2}$ windows are fully momentum-matched (see diagrams for blue and orange band sections). Therefore, increasing both $\Delta\mu_{R2}$ and the fraction of resonant states within it, explains the sharp increase in $J_{ON2\uparrow} = -71, -103$ nA μm⁻² at $V_{R2\uparrow} = -0.71 - 0.78$V, respectively. Finally, the small shoulder observed in all peaks is attributed to a finite $\approx 0.3°$ twist between the BLGs (Fig. S7).

**Tuning the coercive field**

Notably, we find a novel gate-tuning capability of the coercive switching bias: $V_c$, providing a handle to delay the switching dynamics, a feature highly desirable for many device implementations[4–7]. Panel g shows a systematic $V_c$ increase with increasing $V_{gs}$ and $V_R$. While this overall trend is robust, junction-to-junction variations in $V_c$ remain considerable (Fig. S2). This variability is due to solitons' pinning by the rigid moire network[14,48], which remains stochastic in the present junction architecture (SI. Section 2).

Despite these variations, we attribute the underlying solitons' freeze-out mechanism to enhanced screening of $P$ under resonant conditions. As $V_R$ approaches $V_c$, wavefunctions from the source and drain extend, enhancing the electron's probability to reach the polar interface, thereby increasing the dielectric response of the medium. This enhanced screening reduces the polarization magnitude (Fig. 2b, inset) and the shear force on the soliton, thereby increasing the coercive bias.

**Temperature- and bias-dependent polarization**

Under a finite bias $V_{ds}$ and a co-aligned displacement and $P$ orientation, one expects a polarization enhancement (or suppression for anti-aligned orientations). However, at elevated temperature or resonant conditions, we expect depolarization for both $P$ orientations. To track these responses, we analyze the resonant bias difference: $V_{R1\uparrow} - V_{R1\downarrow}$ as a function of carrier density $n_s$, temperature $T$, and $J_{ON}$ magnitude, which reflects the wavefunction overlap. For convenience, we use relatively large multi-domain junctions based on WSe₂ polytypes, with different partial coverages of $P\uparrow$ and $P\downarrow$ domains. Fig.2a shows a Kelvin probe force microscope (KPFM) map of a sample with a typical $\approx 1$ μm² domain area that remains visible outside the top BLG electrode. This relatively large 70 μm² tunneling area (marked by diagonal lines) probes $J$ through both polarization states in parallel within a single scan direction.

Panel b presents forward $V_{ds}$ scans for sample #3 (S3) acquired at 1.5 K for three selected carrier densities, $n_s = 0, \pm 3.75 \times 10^{12} e$/cm² (for more carrier densities, see SI Movie 2). The two resonances observed here for a given scan direction correspond to $V_{R1\uparrow}$ and $V_{R1\downarrow}$ (rather than a single domain peak in Fig. 1).

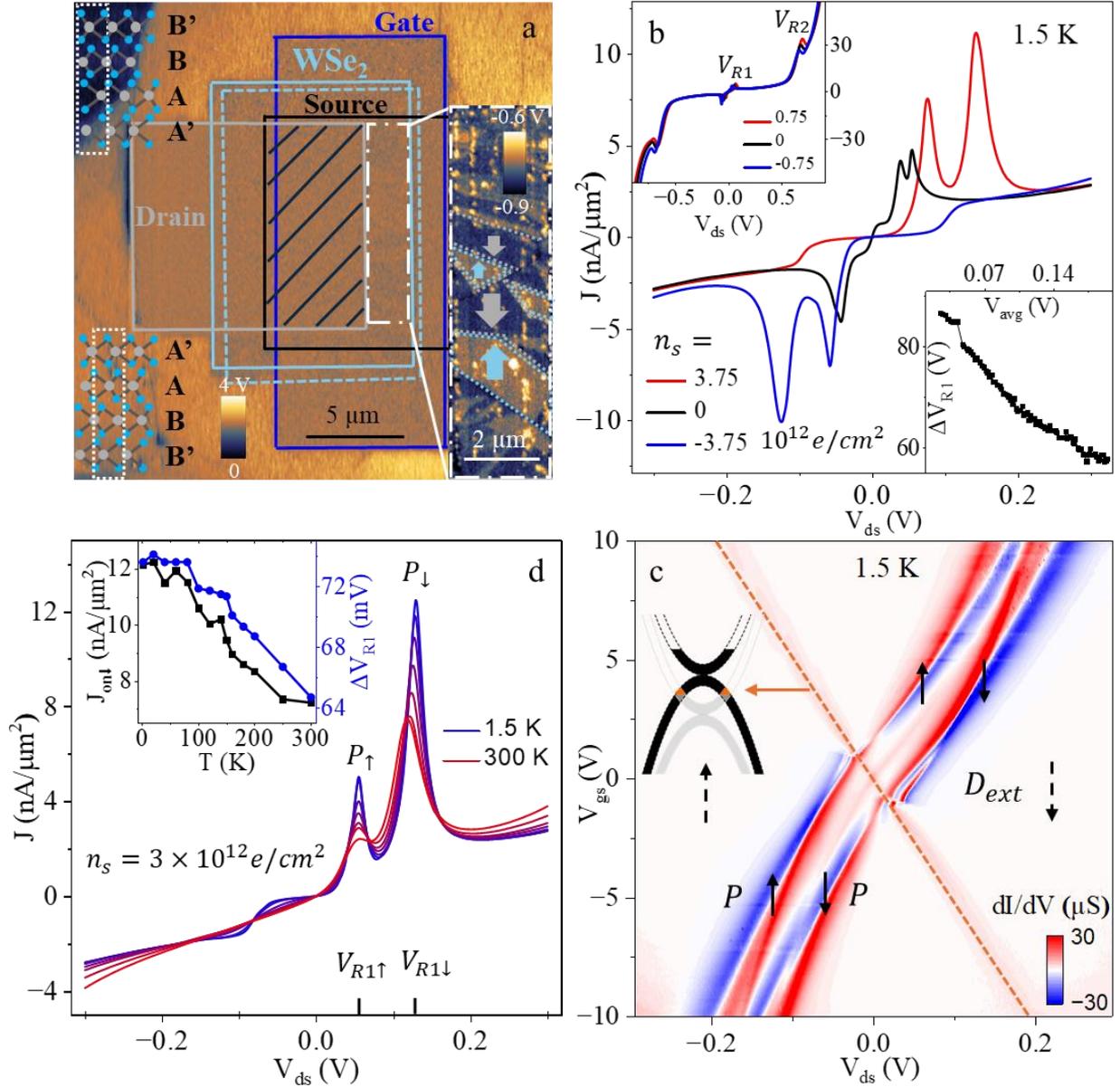

*Figure 2: Temperature and gate-dependent polarization in multi-domain SFeRT junctions of WSe$_2$. (a) KPFM image of S4, showing polar domains outside the top drain BLG with a zoomed image in the right inset. (b) Forward J-V$_{ds}$ curves of S3 for V$_{gs}$ = 0, ± 5 V, and n$_s$ =0, ± 3.75 ×10$^{12}$ e/cm$^2$. Left inset: higher bias scan. Right inset: the distance between the resonant peaks $V_{R1\uparrow} - V_{R1\downarrow}$ as a function of the average $V_{R1}$. (c) differential conductance (dI/dV) map as a function of V$_{ds}$ and V$_{gs}$. Arrows show the orientations of the externally applied (D$_{ext}$) and internal polarization (P). (d) J-V scans at different temperatures from 1.5 to 300K (in 50K steps). Inset: $J_{on\downarrow}$ and $V_{R1\uparrow} - V_{R1\downarrow}$ (right axes) as a function of temperature.*

Additional differences from single-domain junctions are deviations from the ideal $V_{ds}$–P symmetry due to unequal domain coverage, and an exponential (diode-like) enhancement of the current already at $V_{ds} \approx 0.6$ V, close to the second-band resonance $V_{R2} = 0.67\ V$. Panel b, inset, shows $V_{R2}$ shifts with V$_{gs}$ similar to Fig. 1d (see Fig. S3c for extended gate voltages). This behavior reflects the reduced band gap of WSe$_2$ tetralayers (≈1.4 eV), which facilitates Fowler–Nordheim tunneling at lower bias compared to $h$BN (band gap ≈ 5.9 eV),[49] and earlier thermionic injections at elevated temperature (see Fig. S3d).[5]

To trace the $V_R$ evolution as a function of $V_{gs}$ and $V_{ds}$, we plot the differential conductance (dI/dV) map in panel c. In this map, the resonance positions $V_{R1\uparrow}, V_{R1\downarrow}$ appear as two nearly parallel white lines (dI/dV = 0), which converge as $|V_{gs}|$ and $|V_R|$ increase. This convergence is further detected in panel b inset that shows a ≈ 40% drop in the bias separation $(V_{R1\uparrow} - V_{R1\downarrow})$ bewteen the $V_{gs} = 0$ and $V_{gs} = 10$ V curves, corresponding to $\Delta n_s = 7.5 \times 10^{12} e/cm^2$. Such a depolarization mechanism, which is independent of the $P$ orientation, points to enhanced screening by wavefunction overlaps (reflected in a $J_{ON}$ increase) as discussed above. This suppression trend extends to the second resonant peak, where we measure $(V_{R2\uparrow} - V_{R2\downarrow}) \approx 50~mV$ (Fig. SI.3c).

The line running along the transverse diagonal in the map (panel c, dashed line) marks the onset of a $J_{OFF}$ step, which occurs when the band-crossing ring enters $\Delta\mu$ (see the inset band diagram). It remains straight, indicating negligible variations of the dielectric environment within the map parameters range.

The temperature effect on SFeRT is shown in panel d. Thermal broadening of the resonant peaks starts at ≈ 50K, reducing $J_{ON}$ by about half at room T for $V_R = 0.12$ V. Panel d inset also displays a ≈ 15% suppression of $(V_{R1\uparrow} - V_{R1\downarrow})$ at room T reflecting depolarization by thermally-excited carriers in the parent TMD. The thermal response of $\Delta V_R$ and $J_{ON}$ for various $n_s$ is presented in Fig. S3 together with a $J(n_s)$ map. We discuss the partial switching dynamics of the multi-domain junction in Fig. S4, and further results from S4 and S5 (with 6-layer barrier) in Fig. S5,6.

### J-V modeling and predictions for multi-polar junctions

To elucidate the transport and electrostatic mechanisms at play,[20,42] we modeled the device characteristics using a spectral function overlap formalism. The tunneling current density $J$ for a specific polarization branch and $V_{ds}$ is calculated as[50]:

$$I(V_{ds}) \propto \Sigma_{i,l} \int dE \int \frac{d^2k}{(2\pi)^2} A_s^i(k,E) A_d^l(k,E) [f(E - \mu_s, T) - f(E - \mu_d, T)]$$

where $A_{s,d}(k,E) = \frac{1}{\pi} \frac{\Gamma}{(E-E_k)^2 + \Gamma^2}$ are Lorentzian spectral functions representing the source and drain electrodes, broadened by a scattering factor $\Gamma$. This summation accounts for the overlaps between the four Dirac paraboloids from each BLG (16 terms total). The energy dispersion paraboloids $E_k$ are derived from a tight-binding Hamiltonian and integrated into a self-consistent electrostatic framework[51–53] to determine the chemical potentials $\mu_{s,d}$ for each $V_{ds}$, gate voltage $V_{gs}$, and internal polarization $V_P$. To capture the experimentally observed depolarization effects (Fig. 2b, inset), we define the polarization voltage as $V_P = V_P(0) + \beta_\pm V_{ds}$, where $\beta_\pm$ represents a fixed depolarization constant. This approach expands upon previous electrostatic modeling by colleagues for monolayer graphene electrodes[20], developed following our suggestion of the SFeRT concept. We further incorporate Fermi-Dirac occupations at finite temperature, and a momentum mismatch term $\Delta K = \frac{8\pi}{3a}\sin\frac{\theta}{2}$, between the paraboloids to capture finite twist angles $\theta$ between the electrodes. For multi-domain junctions, we introduce a relative coverage parameter $0 < \eta < 1$ to capture the parallel current contributions from both domains. For the full calculations, see SI section 7.

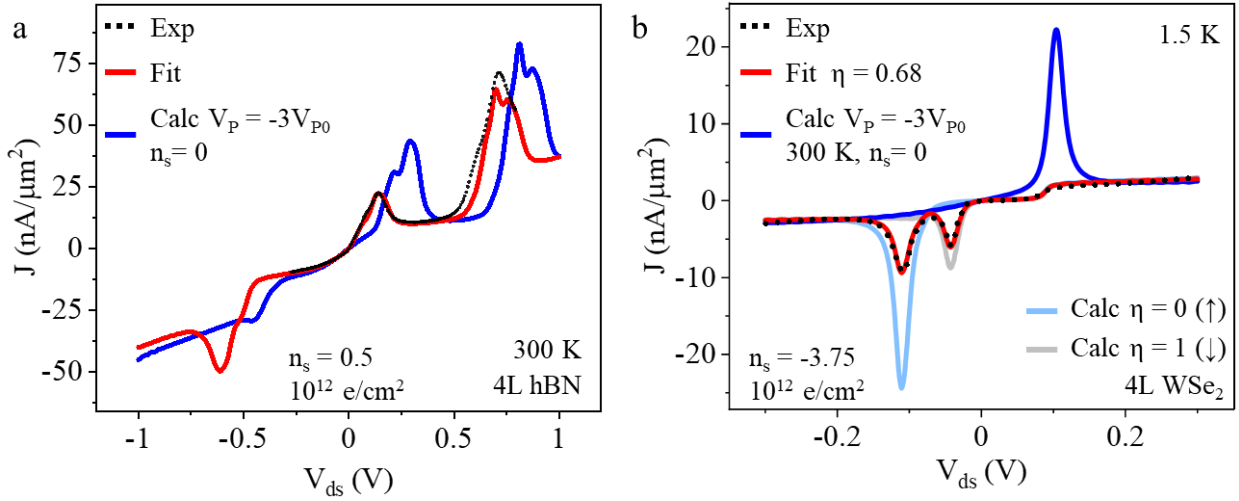

*Figure 3: Simulation and predictions of SFeRT junctions. (a) A simulation fit (solid red) of the J-V curve (dotted black) measured for a tetralayer hBN polytype (A'ABB') with a single polar interface. The solid blue curve is predicted for an ABCA polytype with three co-polar interfaces. (b) A dotted black measurement of WSe$_2$ is fitted (solid red) and used to predict a single-domain junction of A'ABB' at 1.5K, and to predict ABCA polytype of WSe$_2$.*

Fitting results of representative polarizations and temperatures are presented in Fig. 3. Although neglecting indirect tunneling processes (such as Fowler-Nordheim tunneling), we find good agreement with the primary experimental features. For example, see the solid red curves that overlap the experimental dotted black curves of room temperature $h$BN junctions (panel a), and of low temperature WSe$_2$ junctions (panel b). Further calculations and fitting results of $V_{ds}$ - $V_{gs}$ maps are presented in Fig. S7, capturing the central map features, including the gate-dependent depolarization and current magnitudes.

Consequently, we use the dielectric, broadening, and twist parameters obtained from fitting the present junctions to predict $J$-V curves in tetralayer junctions of $h$BN and WSe$_2$ with three polar interfaces; The blue line prediction in panel a for the ABCA $h$BN polytype at room T is calculated using the $\Gamma = 6.8$ meV, $\theta = 0.37°$, $\varepsilon = 3.2$, $\beta = 0.05$ parameters obtained from the solid red fit to the dotted black experimental measurement. It predicts $V_{R1} \approx 0.3$ V and $J_{ON} \approx 50$ nA/µm$^2$ for $n_s = 0$. Similarly, a single-domain junction of fully polarized WSe$_2$ suggests $V_R \approx 0.11$ V and $J_{ON} \approx 20$ nA/µm$^2$ for $n_s = 0$ at room temperature (Fig. 3b, blue line). Notably, these junctions can switch between four total polarization states ($\pm 3V_p$, $\pm 1V_p$) within three distinct crystaline structures ABCA (↑↑↑), ABAB (↑↓↑), and ABCB (↑↑↓), providing a rich $J$-V and multi-ferroic response.

**Conclusions**

We report single-domain SFeRT junctions comprising a single polar interface embedded within tetralayer $h$BN and WSe$_2$ polytypes. At room temperature, these devices exhibit $J_{ON}$ values of approximately 18 and 5 nA per square micrometer at reading biases ($V_R$) of about 0.12 and 0.05 V, respectively, with $J_{OFF}$ nearly fivefold lower. The OFF-state current is primarily determined by residual resonance at band-crossing rings. Application of a gate voltage ($V_{gs}$) linearly tunes the resonant condition, increasing $J_{ON}$ to roughly 45 nA per square micrometer (per $3 \times 10^{12}$ $e$/cm²) doping) without compromising the OFF state. This scaling behavior indicates that increasing the number of polar interfaces - thereby raising $V_P$ - should proportionally enhance the TER without doping (Fig.3). Alternatively, reducing the number of layers should enhance $J_{ON}$ and $J_{OFF}$ by ≈ 20 times per layer, up to $J_{ON} \approx 20$ µA/µm² in polar bilayer polytypes. The layer count, therefore, provides a practical design parameter for balancing the reach switching response - with N distinct polarizain states for N layers (for N-1 polar interfaces) respectively - against absolute ON current. We emphasize the source-drain symmetry of SFeRT junctions, which is unavailable in conventional FTJs, and enable high ON currents in both positive and negative biases.

Higher-order BLG resonances emerging near $V_{R2} \approx 0.7$ V yield larger current densities, but concurrent depolarization reduces the overall TER. Alternatively, monolayer graphene electrodes with lower $C_q$ at $V_{ds} < 0.4$ V enable sharper $V_R$ tuneability. The overall nonlinear, non-monotonic, and non-volatile $P$-dependent $J$-$V$ characteristics establish SFeRT junctions as a promising platform for memory and in-memory computing. The sharp single-point switching characteristics and low coercive voltages ($V_c < 0.5$ V) confirm rapid, durable, and energy-efficient sliding of van der Waals solitons. Notably, the crystalline precision of these heterostructures minimizes device-to-device current variation, addressing a longstanding limitation in scaling non-volatile memristive elements, which rely on stochastic hysteresis mechanisms such as filament formation[7]. Present variations in SFeRT $J_{ON}$ and $V_R$ values are due to finite twist angles between the electrodes (Fig. S7), while $V_c$ variations are due to soliton pinning (Fig. S2). Building on the latter understanding, we expect the SLAP architecture to suppress coercive-voltage variability substantially through engineered superlubric interfaces.

We further identify a gate-tunable control pathway for $V_c$: as $V_R$ approaches $V_c$, resonant depolarization - estimated at ≈ 30% per $\Delta V_{R1} = 0.1$ V (for WSe$_2$) - locally reduces the shear force acting on the soliton, effectively increasing the coercive field. This gate-controlled switching delay may enable selective addressing in high-density crossbar arrays[7].

Together, resonant tunneling and superlubric sliding of solitons position SFeRT junctions as a foundational element of the emerging "slidetronics" paradigm[15]. Beyond memory technologies, their low-bias sensitivity and abrupt switching thresholds (Fig.1c) point to immediate opportunities in mechanical and electro-optical sensing. We foresee devices that switch at $V_c < V_p$ (negative-capacitance) regime, that further combine a negative differential conductance and current instability. Such instabilities should produce excitable dynamics and spiking behavior in SLAP cross-bar arrays for artificial neural networks.

**Acknowledgements**

We acknowledge Y. Lahini, E. Sela and D. Koprivica for discussions, and N. S. Ravid, I. Y. Malker and P. Yanovich for laboratory support. Z.S. was supported by ERC-CZ program (project LL2101) from Ministry of Education Youth and Sports (MEYS) and by the project Advanced Functional Nanorobots (reg. No. CZ.02.1.01/0.0/0.0/15_003/0000444 financed by the EFRR). K.W. and T.T. acknowledge support from the JSPS KAKENHI (grant numbers 21H05233 and 23H02052) and World Premier International Research Center Initiative (WPI), MEXT, Japan. M.B.S. acknowledges funding by the European Research Council under the European Union's Horizon 2024 research and innovation program ('SlideTronics', consolidator grant agreement no. 101126257) and the Israel Science Foundation under grant no. 319/22. We further acknowledge the Centre for Nanoscience and Nanotechnology of Tel Aviv University.

# A Sliding Ferroelectric Resonant Tunnel Junction
# Supplementary Information

**Contents**

**Table of Figures**

## SI.1. SFeRT device fabrication

BLG flakes are exfoliated on Si/SiO$_2$ wafer and pre-patterned into two pieces, which will be used as source and drain electrodes (Fig. S1a), using a pulsed laser etching (NPI Lasers Rainbow OEM 1064, 15 ps pulsewidth). As a spacer layer to incorporate SLAP architecture, we use a ~2 nm thick h-BN flake with cavities, which are patterned by Electron beam lithography (EBL), followed by CHF$_3$:O$_2$ plasma etching [1,2]. Bilayer h-BN and Bilayer WSe$_2$ flakes are identified optically and cut using AFM diamond tip (MikroMasch HQ:DMD-XSC11) while applying high force in contact mode (see Fig. S1d for Bilayer WSe$_2$ after AFM cut).

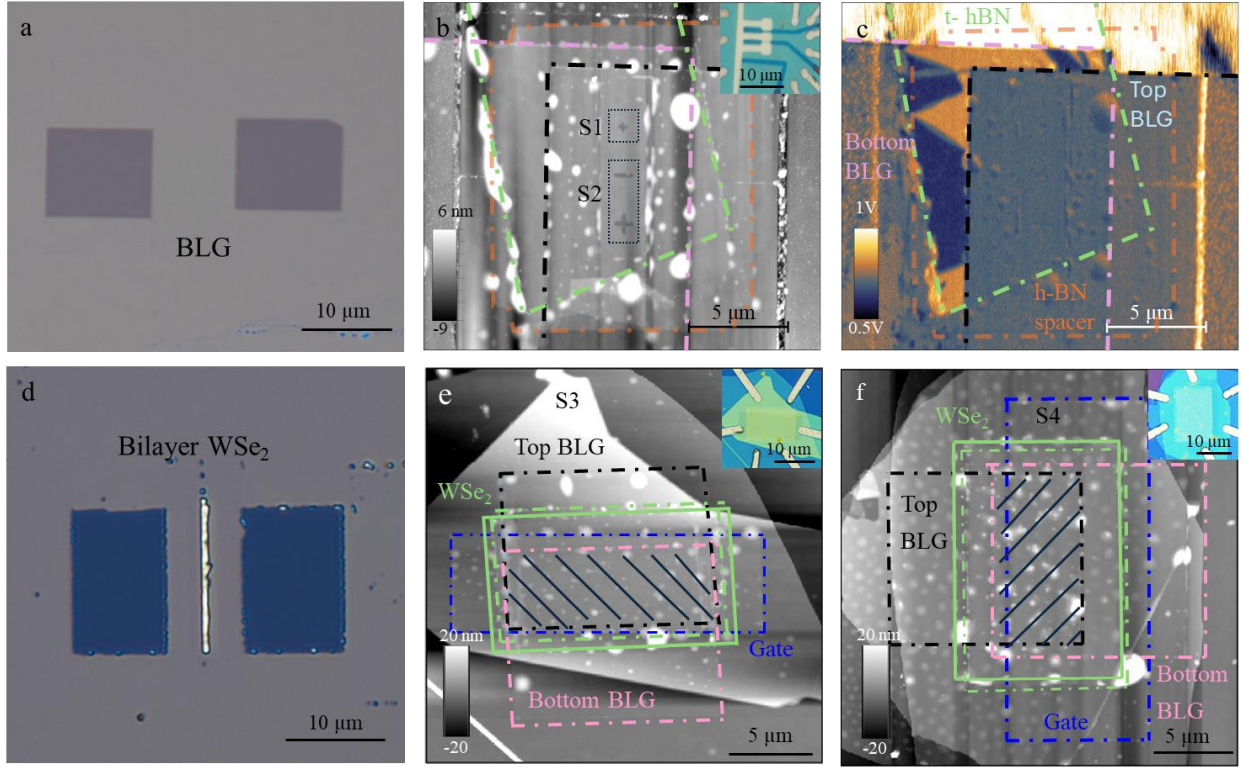

*Fig. S 1: SFeRT device fabrication.* (a) Optical image of BLG after pulsed laser etch. Topography (b) and KPFM (c) micrographs of h-BN SFeRT junction for samples S1 and S2 prior to metallization. S2 consists of two merged cavities due to the unification of two gates. (d) Optical image of Bilayer WSe$_2$ after designed cutting through AFM. (e, f) Topography micrographs of Samples S3 and S4, respectively. The shaded region is the device's active area. Insets, optical images of the devices after metallization.

The h-BN SFeRT heterostructure is built top down with PDMS/PMMA stamp in the following pickup sequence: h-BN, first BLG, first bilayer h-BN, second bilayer h-BN, h-BN spacer, second BLG. Finally, the stack is dropped on a large hBN flake. For WSe$_2$ SFeRT: WSe$_2$ were made by chemical vapor transport in quartz ampoule directly from elements using bromine as a transport medium. Tungsten (99.999%, -100 mesh, China Rhenium Co, China) and selenium (99.9999%, 2-4 mm, Wuhan Xinrong New Materials Co, China) were placed in stochiometric ration corresponding to 50 g of WSe2 in quartz ampoule (50x250 mm) together with 2 at.% excess of selenium and 1 g of SeBr4 (99.9%, STREM, USA) and melt sealed under high vacuum (<1x10-3 Pa using oil diffusion pump and LN2 cold trap). The ampoule was first heated in muffle furnace on 500 °C for 25 hours, on 600 °C for 50 hours and on 800 °C for 50 hours. Heating rate were 1 °C/min and free cooling. Subsequently the ampoule was placed in two zone furnace for crystal growth. First the growth zone was heated on 1000 °C and source zone on 800 °C. After 2 days is thermal gradient reversed and source zone is heated on 1000 °C and growth zone on 900 °C for 14 days. Finally, the ampoule is cooled on room temperature with growth zone additional heated in 300 °C for 2 hours to remove volatile compounds. Finally, ampoule was open in argon glovebox and crystals collected.

The WSe$_2$ SFeRT stacking sequence is as follows: h-BN, first BLG, first bilayer WSe$_2$, second bilayer WSe$_2$, second BLG, h-BN dielectric, few-layer graphene gate, large bottom h-BN flake. The two BLG flakes are picked up in the same orientation using a motorized stage, and the stage is rotated by 180º between the pickup of the first and second bilayer h-BN or WSe$_2$ to create a parallel-stacked (AB/BA) polar interface. The h-BN fabricated stack is then cleaned by AFM in contact mode because the cavities tend to accumulate contamination. Contacts are then patterned using a standard EBL procedure, and Cr/Nb/Pt 5/45/15 nm contacts are sputtered using AJA Orion-8 system. Contacts are made to BLG by etching top h-BN using CHF$_3$:O$_2$ plasma, followed by etching of BLG by O$_2$ plasma. Top gates in h-BN devices, above the cavity region, are then patterned similarly. To measure each cavity individually, the top gates are connected outside of the top BLG electrode (inset Fig. S1b). Topography and KPFM micrographs of hBN SFeRT junctions (for S1, S2) are shown in Fig S1b and S1c, respectively. Topography images of WSe$_2$ SFeRT junction for samples S3 and S4 are shown in Fig. S1e and S1f, respectively. The insets show the optical image of the device after contacts.

Table of device parameters:

|  | Tunnel barrier | Device area | Gate dielectric thickness | Single/multi domain |
|---|---|---|---|---|
| Device 1 | 4L h-BN | 0.08 µm$^2$ | 22 nm | Single |
| Device 2 | 4L h-BN | 0.35 µm$^2$ | 22 nm | Single |
| Device 3 | 4L WSe$_2$ | 70 µm$^2$ | 24 nm | Multi |
| Device 4 | 4L WSe$_2$ | 70 µm$^2$ | 10 nm | Multi |
| Device 5 | 6L WSe$_2$ | 50 µm$^2$ | 30 nm | Multi |

## SI.2. Extended J-V measurements

The main text presents measurements of 2 hBN junctions and a WSe$_2$ junction selected out of more than 15 devices studied overall, owing to small twist angles between the electrodes. Table S1 specifies three more devices for which we present measurements ahead.

**Tunneling current density variations**

We attribute the main junction-to-junction variations in $J$ to variations in the twist angle $\theta$ between the electrodes. Our simulations suggest a 10% suppression in $J$ for $\theta$=0.065 degrees and a $\Delta V_{R1} = 16\ mV$ per 0.1 degree for BLG electrodes and hBN barrier at room temperature (Fig. S7d), in agreement with previous studies of resonant tunneling between marginally twisted electrodes through non-polar polytypes[3]. We note that for twist angles below ~ 0.05 degrees, the momentum shift is smeared by broadening, and the $J_{ON}$ response of different junctions becomes uniform, regardless of the junction dimensions – see Fig. 1a, b.

**Impact of twist between the dielectric layers**

A marginal twist angle $\varphi < 0.5$ between the dielectric layers does not change the magnitude of $P$.[4,5] Rather, it affects the switching dynamics as was previously studied in detail.[6,7] The larger the twist, the stiffer the soliton network becomes, and the switching is suppressed. Increasing $\varphi$ from 0 (only tensile solitons) first increases $V_c$, then makes the system paraelectric-like without hysteresis, and finally freezes out the soliton and prevents switching.

In the single-domain junction architecture, junction diameters are shrunk significantly below the distance between the solitons.[1,8] These samples are at the $\varphi \ll 0.1$ limit, where $V_c$ is likely determined by residual soliton pinning from stochastic edge disorder rather than the moire stiffness. We believe that S1 and S2 express the switching dynamics and pinning potential of a single soliton.

In S1, switching events occur in a repeatable and reliable manner, and the cohesive fields shift under applied gate-source voltage. This behavior is demonstrated clearly in the forward scan presented in Fig. S2c, within the dashed region, where sharp jumps in conductance (switching events) drift to higher $V_{ds}$ under higher gate values. In S2, however, the switching in the forward direction is stochastic (dashed region in Fig. S2e), and there is no clear relation between the cohesive field and gate voltage. Moreover, at a gate voltage of 2V and above, switching cannot be achieved within $V_{ds} = \pm 0.8V$, and the system becomes limited to one polarization state only (↑ state, dashed horizontal line in Fig. S2e, S2f). This behavior agrees with enhanced polarization screening close to resonant conditions as discussed in the main text.

While we find the switching response and $V_c$ values appealing for many applications, we expect a much better control and uniformity in SLAP junctions, based on our previous work that studied the switching dynamics of more than 1000 polytype junctions[8].

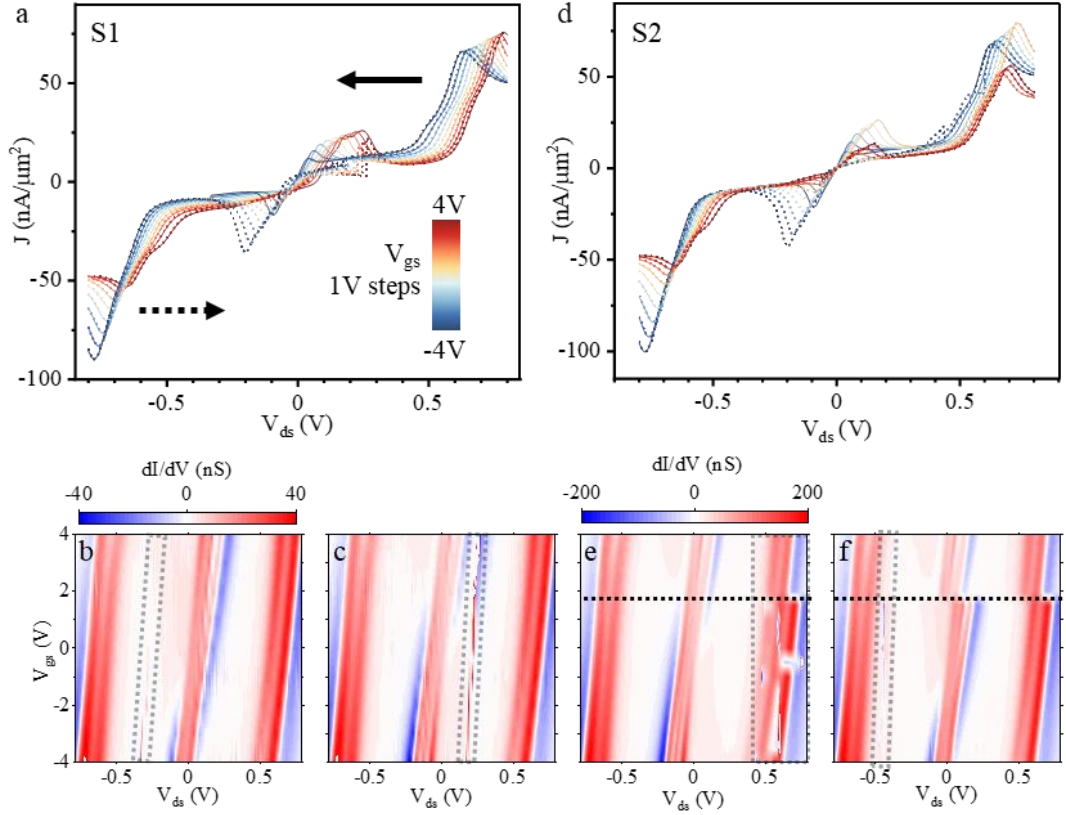

*Fig. S 2: Extended data for single domain switching SFeRT. J-V response (**a,d**) and conductance (dI/dV) of backward (**b,e**) and forward (**c,f**) scans for S1 (a-c) and S2 (d-f). Polarization switching regions are indicated by gray dashed boxes. For $V_g > 2$ V (dashed horizontal line), Sample 2 remains in the up polarization without switching.*

## SI. 3. Extended measurement of sample S3

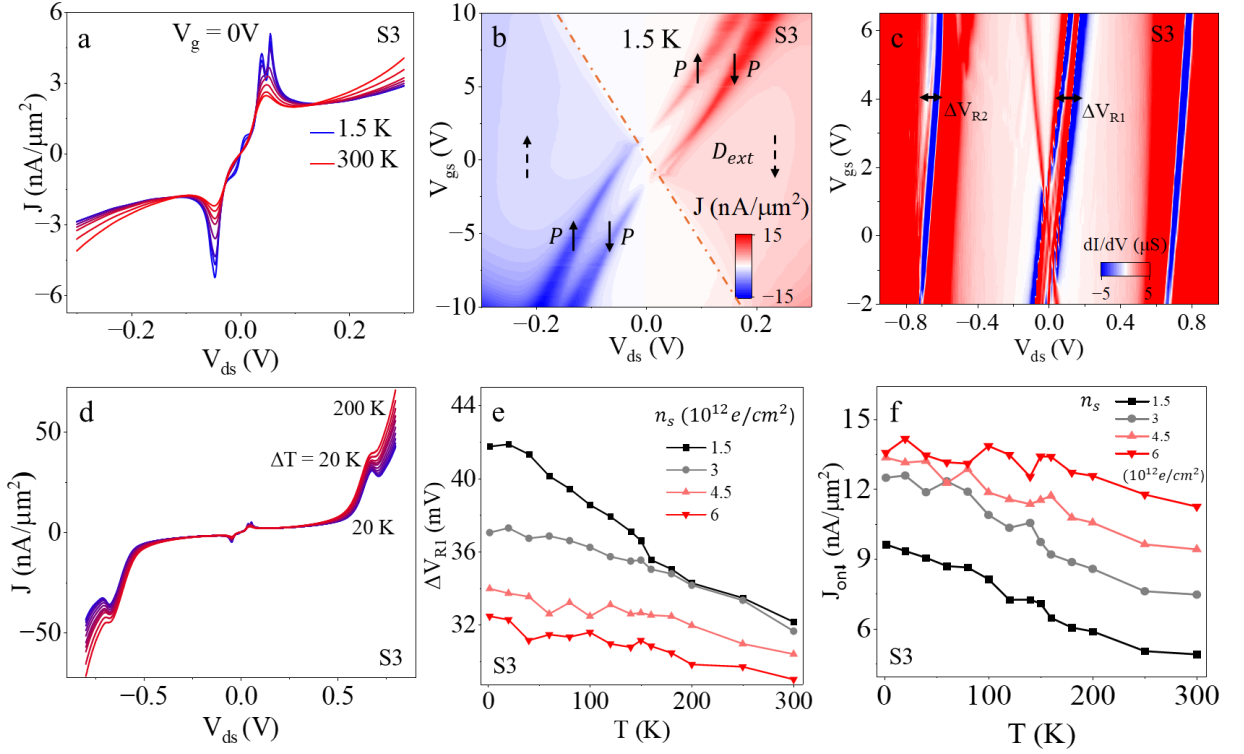

*Fig. S 3: Extended data of sample S3. (a): J-V response at $V_g = 0V$ for different temperatures (in steps of 50 K). The observed splitting at low temperature may originate from twist, strain, or trigonal warping of electronic dispersion in BLG. (b): Current density color plot as a function of $V_{ds}$ and $V_g$. (c): conductance (dI/dV) color map as a function of $V_{ds}$ and $V_g$. $\Delta V_{R1}$ and $\Delta V_{R2}$ represent the difference in up and down polarization for the first and second resonance peaks. (d) Temperature-dependent J-V response at high drain bias. (e,f): $\Delta V_{R1}$ and $J_{ON}$ for down polarization as a function of temperature for different carrier densities.*

## SI. 4. Partial switching in multi domain SFeRT device

The two stable domain configurations ($P_\uparrow$ and $P_\downarrow$) formed by parallel stacking of WSe$_2$ manifest as two resonance peaks in the current density voltage sweep (Fig. S4). The switching of these domains can be achieved by applying a perpendicular electric field. The positive electric field aligns with down polarization and anti-aligns with up polarization, making the down polarization area more stable after it crosses a coercive potential of $0.18\ V$ (Fig. S4a). Similarly, the negative electric field aligns with the up polarization and anti-aligns with down polarization, making the up-polarization area more stable after it crosses a coercive potential of $-0.17\ V$ (Fig. S4b). These results agree with the literature reports on switching dynamics with polar domains.[9,10] The change in the current density amplitude for up and down polarization indicates the motion of domain boundaries with electric field. See Movie 2 for the hysteresis in current voltage plot for different gate voltages. While the large area devices exhibit multi domain structure which leads to partial switching, reducing the device footprint can enable a complete switch between up and down polarization as in case of hBN SFeRT.

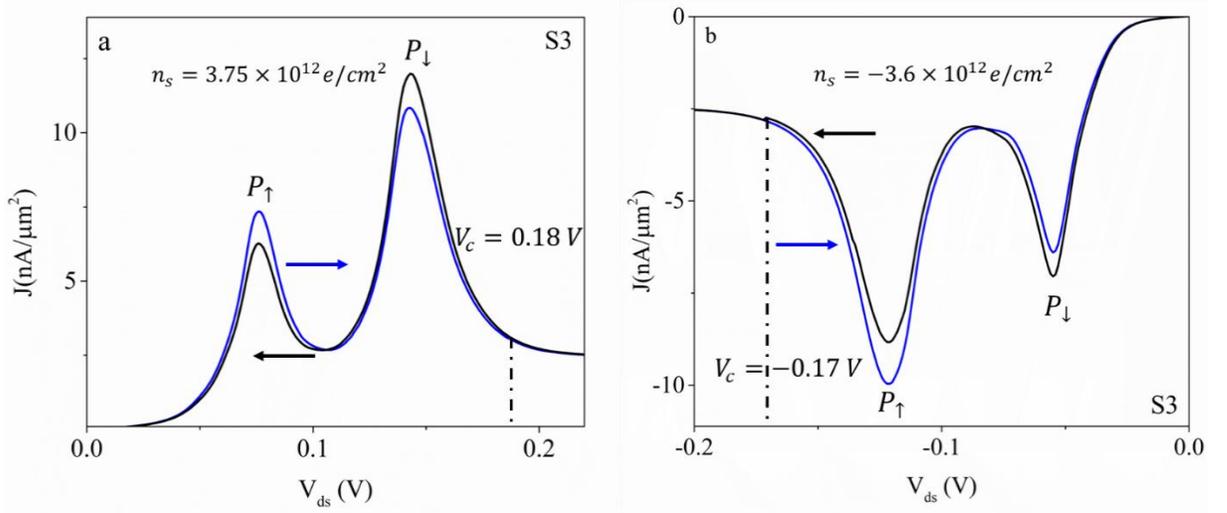

*Fig. S 4: Hysteresis current density-voltage response.* J-$V_{ds}$ plot of WSe$_2$ SFeRT S3 device for $n_s = 3.75 \times 10^{12} e/cm^2$ (a) and $n_s = -3.6 \times 10^{12} e/cm^2$ (b). Blue and black arrows indicate the forward and backward sweep directions.

**SI. 5.** Measurement of S4 – J-V response, T and $V_{gs}$ dependent P

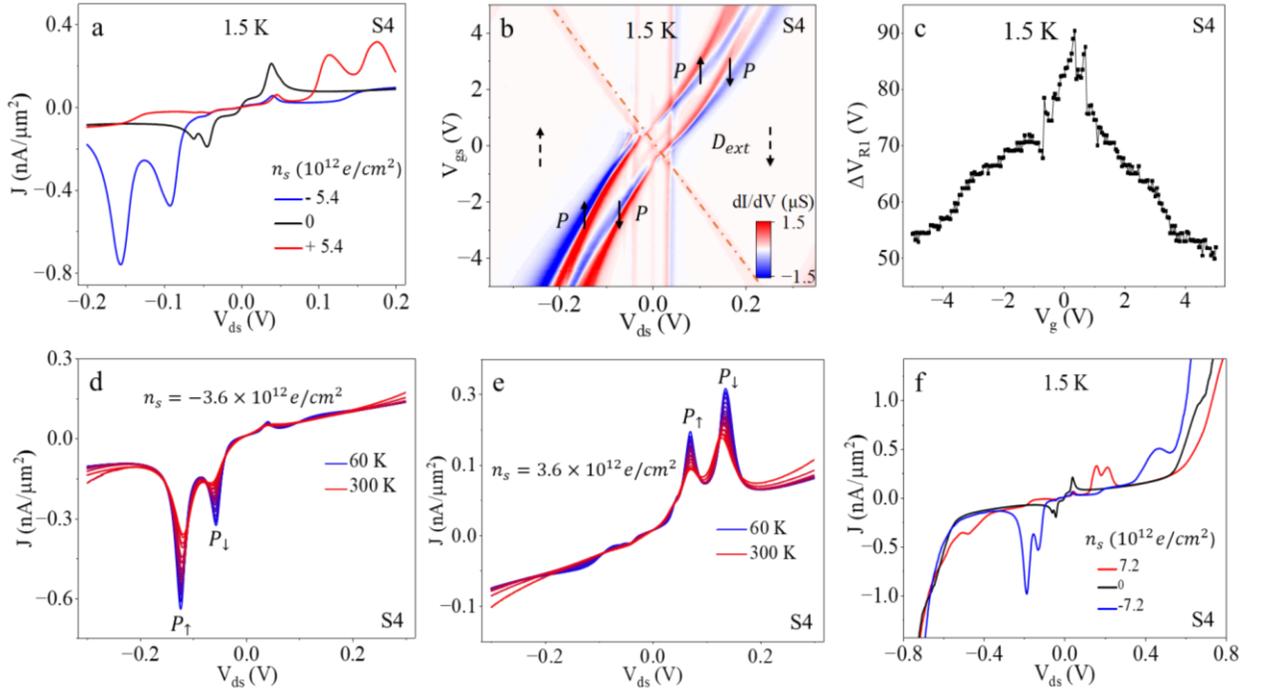

*Fig. S 5: Current-voltage response of sample S4.* (a): J-V response of WSe$_2$ device. (b): conductance color map as a function of $V_{ds}$ and $V_g$. (c): Difference between up and down polarization as a function of gate voltage. (d, e): Temperature dependent J-V for $n_s =$

$-3 \times 10^{12} e/cm^2$ and $n_s = 3 \times 10^{12} e/cm^2$ respectively. (f): High bias J-V response with second resonance peak.

The sample S4 has the same device geometry as S3, with a different proportion of polarization as reflected by asymmetric J-V response (see Fig. S5a). We find similar gate response – see the maps in Fig. 2c and Fig. S5b for S3 and S4, respectively, and a similar depolarization effect from 90 $mV$ at $V_g$= 0 to 50 $mV$ for $V_g$= 5 V (Fig. S5c for S4 and Fig. 2b right inset for S3). Temperature-dependent thermal broadening of peaks for negative and positive current density is indicated in Fig. S5d and S5e respectively. The higher drain bias features the second resonance peak at 0.46 V (for gate voltage of –4 V) and – 0.47 V (for gate voltage of 4 V) (Fig. S5f).

## SI. 6. Measurement of Sample S5

Sample S5 consists of 6L WSe$_2$ as a sliding ferroelectric tunnel junction between BLG. The device features a similar response to sample S3 in terms of bias and gate voltage dependence (see Fig. S6a and S6b for sample S5 and Fig. S3 for sample S3) except for the magnitude of current. Current density is reduced by 2 orders of magnitude with the addition of 2L WSe$_2$. The gate dependent difference of the peak position of the polarization states, $\Delta V_{R1}$, reduces from 80 mV at $V_{gs} = 0$, to 44 mV at $V_{gs} = 8V$, which corresponds to a change in carrier density $\Delta n_s = 5 \times 10^{12} e/cm^2$ for a 30 nm thick gate dielectric (Fig. S6c).

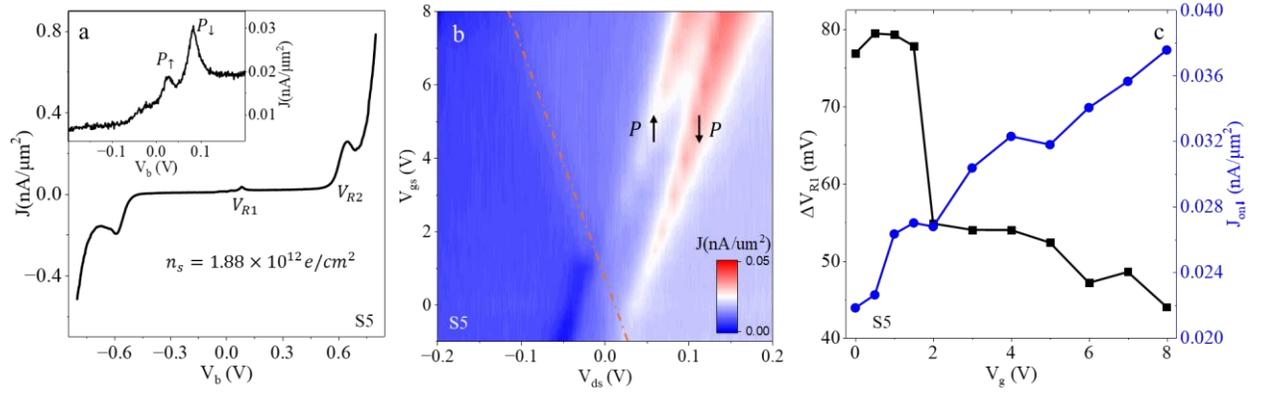

*Fig. S 6: Current voltage response of sample S5.* (a): J-V response of 6L WSe$_2$ SFeRT device. Inset: low bias current density response. (c): Current density color map as a function of $V_{gs}$ and $V_{ds}$. (c): $\Delta V_{RI}$ and $J_{on\downarrow}$ as a function of gate voltage.

## SI. 7. Theoretical calculations of I-V in SFeRT junctions

In the following, we compare our measurements to simulated J-V curves produced using the spectral function overlap model. For bilayer graphene, four energy bands are considered, namely, a pair of conduction and a pair of valence bands. For the calculation of the dispersion relations, we diagonalize a tight-binding Hamiltonian parametrized by $\gamma_{0-4}$, near the K-point[11–13]. The calculated energy bands are fed into a self-consistent electrostatic calculation. For a given temperature $T$ and chemical potential $\mu$, we find the net carrier density per area, given by $n_{net}(T,\mu) = n_c(T,\mu) - p_v(T,\mu)$, Where $n_c(T,\mu)$ is the electron density in the conduction band and $p_v(T,\mu)$ is the density of holes in the valence band. With twofold spin and valley degeneracies, we write, $n_c(T,\mu) = \frac{g_s g_v}{(2\pi)^2} \int d^2k[f(E_{c1}) + f(E_{c2})]$ and $p_v(T,\mu) = \frac{g_s g_v}{(2\pi)^2} \int d^2k[(1-f(E_{v1})) + (1-f(E_{v2}))]$, where $f(E)$ is the Fermi-Dirac distribution. The sheet charge is then $\sigma = -en_{net}(T, \mu)$. We then build the inverse map $\mu = \mu(\sigma, T)$.

At each drain-source and gate voltage pair $(V_{ds}, V_{gs})$, we solve electrostatic equations for the two unknown voltage drops: $u$, the voltage drop upon the ferroelectric layer, and $u_g$, the gate voltage drop. We consider two capacitors, $C_{FE} = \frac{\epsilon_0 \epsilon_{r,FE}}{d_{FE}}$, $C_g = \frac{\epsilon_0 \epsilon_{r,g}}{d_g}$, the capacitance of the ferroelectric layer, and the gate capacitance, respectively. We then have the drain sheet, gate electrode and source sheet charges $\sigma_d = C_{FE} u$, $\sigma_g = C_g u_g$, $\sigma_s = \sigma_g - \sigma_d$. We include an FE-induced built-in potential, given by $V_p(V_{ds}) = \pm V_{FE,eff}(V_{ds})$, where $V_{FE,eff}(V_{ds}) = V_{FE,base} + \beta_\pm V_{ds}$, with $\beta_\pm$ a linear depolarization constant. We define the Fermi levels of the two BLG sheets, $\tilde{\mu}_d = \frac{\mu(\sigma_d, T)}{e}$, $\tilde{\mu}_s = \frac{\mu(\sigma_s, T)}{e}$. The electrostatic equations are[14]

$$u - V_{ds} - V_p - (\tilde{\mu}_d - \tilde{\mu}_s) = 0,$$

$$u_g - V_{gs} - \tilde{\mu}_s = 0.$$

We generalize to finite temperature by replacing the analytic MLG relation between the Fermi energy and charge by a numerically inverted finite temperature BLG Fermi energy as function of charge and temperature, computed from Fermi-Dirac occupations of the four BLG bands. Using the self-consistent solution of the equations, we define $\tilde{V}_p = V_p - u$, and apply symmetric energy shifts to the two electrodes $E_{D,d} = +\frac{e\tilde{V}_p}{2}$, $E_{D,s} = -\frac{e\tilde{V}_p}{2}$ due to a constant DOS. This defines the electrode electrochemical potentials $\mu_d \to \mu_d + E_{D,d}$, $\mu_s \to \mu_s + E_{D,s}$. We implement a twist angle between the two BLG's by shifting and rotating the bottom momentum $k$. The mismatch magnitude is $\Delta K = 2K_D \sin\left(\frac{\theta}{2}\right)$, where $K_D = \frac{4\pi}{3a}$ with lattice constant $a$[15]. For a given polarization branch, we calculate the tunneling current[16,17]:

$$I \propto \int dE [f_B(E) - f_T(E)] \int \frac{d^2k}{(2\pi)^2} \Sigma_{a,b} A_d^a(k, E) A_s^b(k, E),$$

where a,b are the energy bands $(c1, c2, v1, v2)$ and $A$ are Lorentzian spectral functions with broadening $\Gamma$, $A(k, E) = \frac{1}{\pi} \frac{\Gamma}{(E - E_k)^2 + \Gamma^2}$. We write the total current as, $I_{tot} = \eta I_\downarrow + (1 - \eta) I_\uparrow$. With $\alpha \in [0,1]$. For sharp polarization switching we use $\alpha = 0,1$ for the two distinct currents corresponding to the two polarizations. In the more general case where polytypes coexist we use any $\alpha$ between 0 and 1.

.

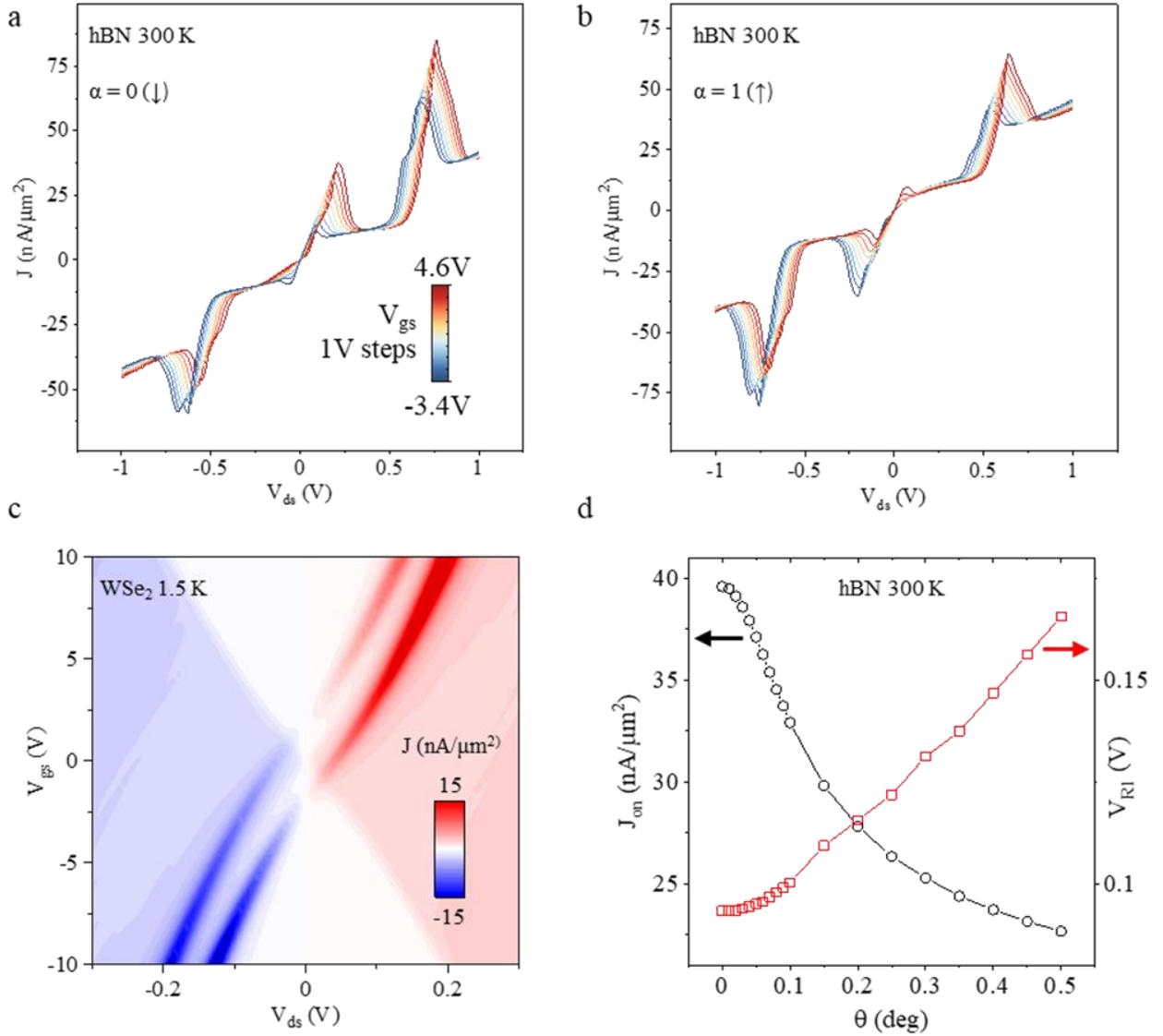

*Fig. S 7: Theoretical calculations of SFeRT junctions.* Calculated gate dependent J-V response of hBN SFeRT junctions S1 and S2 for down (**a**) and up (**b**) polarization states. (**c**) Calculated current density map of WSe$_2$ device S3. (**d**) $V_{R1}$ peak current density (left panel) and position (right panel) dependence on graphene electrodes' misalignment angle. (**e**) Summary table of parameter used in theoretical calculations.